*Article*

# Development, Processing and Applications of a UV-Curable Polymer with Surface Active Thiol Groups

**Manuel Müller \*, Rukan Nasri, Lars Tiemann and Irene Fernandez-Cuesta \***

Center for Hybrid Nanostructures (CHyN), Institut für Nanostruktur- und Festkörperphysik (INF), Universität Hamburg, Luruper Chaussee 149, Hamburg, 22761, Germany; rukan.nasri@physnet.uni-hamburg.de (R.N.); lars.tiemann@physik.uni-hamburg.de (L.T.)

\* Correspondence: mamuelle@physnet.uni-hamburg.de (M.M.); ifernand@physnet.uni-hamburg.de (I.F.-C.)

**Abstract:** We present here a novel resist formulation with active thiol groups at the surface. The material is UV curable, and can be patterned at the micro- and nanoscale by UV nanoimprint lithography. The resist formulation development, its processing, patterning and surface characterization are presented here. In addition, a possible application, including its use to modify the electrical properties of graphene devices is shown. The cured material is highly transparent, intrinsically hydrophilic and can be made more hydrophilic following a UV-ozone or an $O_2$ plasma activation. We evaluated the hydrophilicity of the polymer for different polymer formulations and curing conditions. In addition, a protocol for patterning of the polymer in the micro and nanoscale by nanoimprinting is given and preliminary etching rates together with the polymer selectivity are measured. The main characteristic and unique advantage of the polymer is that it has thiol functional groups at the surface and in the bulk after curing. These groups allow for direct surface modifications with thiol-based chemistry e.g., thiol-ene reactions. We prove the presence of the thiol groups by Raman spectroscopy and perform a thiol-ene reaction to show the potential of the easy "click chemistry". This opens the way for very straightforward surface chemistry on nanoimprinted polymer samples. Furthermore, we show how the polymer improves the electrical properties of a graphene field effect transistor, allowing for optimal performance at ambient conditions.

## 1. Introduction

Nanoimprinting lithography (NIL) has emerged over the last two decades as a high throughput nanofabrication method and is now a consolidated technology [1–3]. The method uses a stamp, brought into physical contact with a polymer, to mold it and transfer the pattern of the stamp into the polymer. Several different approaches can be found for nanoimprinting, being the main ones thermal NIL and UV NIL. In recent years a strong industrially-driven research can be observed, pushing NIL and its applications from purely academic purposes to innovations in data storage, point of care diagnostics, electronic and graphene devices or augmented reality [4–9]. With this shift, new materials need to be developed with properties that are compatible with mass production: fast curing and imprinting, robustness, reproducibility, and patterning on flexible substrates, both, at the micro and nano scale.

NIL was initially used as a purely lithographic method, for pattern transfer [10] as a fast and cheap alternative to electron beam lithography. In recent years, a new trend has emerged towards direct patterning of functional materials, where the imprinted structures are directly used as active elements in devices. These devices find applications in several different fields, like micro and nanofluidics [11,12], biosensing [13] and DNA analysis [14], optical, nanophotonic [15] and flexible, electrical devices [16]. For example, biocompatible and three dimensional polymer structuring for cell cultivation open a promising field for imprintable polymers as an alternative to conventional processing using photoresists, which need long process times, by offering more flexibility in the processing, higher throughput and incrementing the design freedom [17].

For using patterned surfaces in devices or sensors, the surface of the cured polymer often needs to be chemically modified or activated. The functionalization endows the surface with properties different and independent from the bulk or from other untreated areas of the polymer. However, the process of functionalization often requires an activation, or a two-step modification of the surfaces with strong (and dangerous) chemicals or physical interaction [18,19]. The process is time consuming and not always compatible with the device (e.g. compatibility of biomolecules with temperature changes, chemicals or pH changes). Functional, imprintable polymers with intrinsic and easily modifiable functional groups are very attractive alternatives.

A versatile functional group for surface modifications is the thiol group due to the well-researched "click chemistry" [20] and also its strong affinity to common materials used in micro and nanoengineering, such as gold [21]. Thiols are known to undergo a nucleophilic addition with Michael systems, such as (meth)acrylates or vinyls, at room temperature [22–24], as well as radical additions. Spatially controlled reactions by ink-jet printing or masking could be used for local chemical modification of surfaces and patterns on an imprinted structure.

In this work, we have developed a resist formulation, which has functional thiol groups at the surface and in the bulk. It is easily patternable by UV NIL, both at the micro and nanoscale. In addition, it is possible to etch it selectively and homogeneously by reactive ion etching (RIE) using an oxygen plasma. The surface properties have been characterized by measuring the contact angle for different formulations and curing conditions. The thiol groups have been observed by Raman spectroscopy, confirming their presence at the surface and in the bulk, and also after an $O_2$ etching. As a first application, we show the possibilities for surface click chemistry on the polymer surface. And, as a second possible application, we show how the polymer, just placed on top of a graphene device, changes its electrical properties and allows for operation at ambient conditions, without ultra-high vacuum.

## 2. Methods

### 2.1. Curing of the Material

Some of the results shown in this work were obtained by curing the polymer with a monochromatic light emitting diode (LED) lamp (365 nm) under a home-made $CO_2$ flow box. These include those shown in Section 3.2, Section 4.2.2 and in 4.3.2 (nanometric lines). The rest of the results were obtained using a dedicated UV-NIL equipment (EVG 501, EV Group, Sankt Florian am Inn, Austria), with a controlled atmosphere in the chamber, and a pulsed, broadband xenon lamp (Model LH-810 Spiral Lamp 107-mm, type C spectrum, 190 nm cutoff, power 105 mW/cm$^2$, XENON Corporation, Wilmington, MA, USA) for exposure.

### 2.2. Contact Angle

The time-dependent water contact angle (WCA) measurements presented in Figure 2 were done in a Krüss DSA25 (Drop Shape Analyzer) (Krüss GmbH, Hamburg, Germany) in the sessile drop configuration. It should be noted that the dynamic contact angle in Figure 3 differs significantly from the sessile drop configuration, resulting in comparably smaller contact angles.

### 2.3. Dynamic Contact Angle

The setup to measure the dynamic water contact angle of the measurements presented in Figure 1 is a self-made device with a video recording system and a manual syringe for solvent application. A drop of water is placed on the flat cured surface of one of the prototypes. A video is recorded while the water droplet volume increases by slowly pressing out the water out of the syringe. The recorded video is then analysed image by image and the WCA is calculated assuming a perfectly spherical shape. For each droplet, the WCA was calculated seven times along the growth of the droplet. On each surface three different measurements were conducted and the average over all was used to compare the different surfaces.

*2.4. SEM*

SEM images were obtained in a Crossbeam 550L system from Zeiss (Oberkochen, Germany). A low acceleration voltage (4 kV) was used, to avoid charging of the polymers and for an optimal image quality. Some of the images were obtained with the sample tilted at 30°.

*2.5. Raman Spectroscopy*

The measurements are performed in a self-built setup consisting of a Ti-U inverted microscope (Nikon, Tokio, Japan) with a dichroic mirror > 538.9 nm, a 100x objective with NA = 0.90, a DPSS green laser (532 nm, 35 mW at focal point) and an Andor spectrometer (1200 l/mm grating, iDus 420 CCD detector, Andor, Belfast, United Kingdom ). A confocal pinhole (105 μm) is used to reduce the out-of-focal-plane signal.

*2.6. Surface Modification for Contact Angle Measurements*

The UV-ozone surface activation was done in a UV-ozone Cleaner 144AX-220, from Jelight Co Inc. (Irvine, CA, USA). The $O_2$ plasma treatment was done in an oven (Technics Plasma 100-E, 2.45 GhZ, Technics Plasma GmbH, Kirchheim bei München, Germany) at 180 W, 2 mbar, with $O_2$ flow.

*2.7. Graphene Transfer and Graphene Field Effect Transistor (GFET) Sample Fabrication*

Monolayer graphene synthesized by chemical vapour deposition (CVD) on copper foil was released from its growth medium, cleaned and transferred to a *p*-doped Si substrate with a 100 nm surface layer of $SiO_2$ for electrical insulation. A 200 μm by 20 μm Hall bar structure was patterned by standard optical lithography, $O_2$ plasma etching and PVD of Au for the Ohmic contacts. A detailed description of the fabrication process can be found in [25]. The device was then transferred to a ceramic chip carrier and wire-bonded using a combination of ultrasonic wedge bonding (for the chip carrier pads) and conductive silver paint (for the graphene contacts). The *p*-doped substrate was also connected and acted as the back gate. Suitably sized polymer flakes were peeled off of the fabrication substrate, carefully oriented and placed over the graphene Hall bar under a microscope using a pair of tweezers. The flakes were fabricated by curing in contact with a non-polar, fluorinated stamp in the EVG 501, with a pulsed, broadband UV xenon lamp with 105 mW/cm$^2$ for 19 seconds with 2000 mJ/cm$^2$ total dose.

*2.8. Electrical Measurements*

We employed a standard lock-in method for the electrical measurements. The chip carrier with the sample was connected to a probe station and remained exposed to atmosphere/ambient air at room temperature during the measurements. The internal oscillator of a SR830 lock-in amplifier (Stanford Research Systems, Sunnyvale, CA, USA) was connected to a 100 Mega-Ohm resistor in series to the sample to generate a constant low frequency (ac) current of 2 nA (13.33 Hz) in the Hall bar. Changes in the conductivity were measured with the lock-in's differential amplifier between opposite contact pairs where the graphene was either covered by the polymer or exposed to ambient conditions. These contact pairs also acted as source and drain contacts (two-terminal measurements). The gate voltage, applied with a source measurement unit (2400 SMU, Keithley Cleveland, OH, USA), was swept in steps of 100 mV with an integration time of 1 second.

## 3. Development of the Material Formulation

The goal of this work was to develop a resist formulation capable of fulfilling industrial needs for rapid prototyping of micro and nanostructures for with hydrophilic surface properties, while incorporating chemical functionality into the final device. Balancing the ratio of different functional groups in the formulation, a material, which is further capable of post-cure functionalization via thiol-ene click chemistry on the surface, was obtained. Patterning of the functional polymer is straight-forward by direct curing by UV light, as sketched in Figure 1.

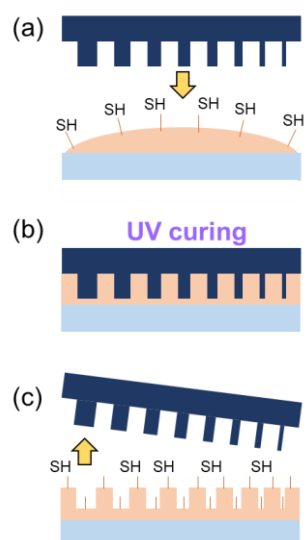

**Figure 1.** Micro and nanostructuring of the thiol polymer. The polymer (orange) with thiol-groups (-SH) intrinsic to its formulation is drop casted or spin coated on the substrate (blue) (**a**). By bringing in contact with a stamp (dark blue), the polymer fills the structure and is cured with UV light (**b**). After curing, the stamp is released and the polymer surface has the positive structure of the stamp and the functional thiol group (**c**).

*3.1. Base Formulation*

When developing a new UV-curable resist, different material characteristics need to be addressed and taken into account, such as nature of the cross linking, curing speed, hardness, flexibility, oxygen insensitivity during curing, hydrophobicity and many more. By choosing the right monomers, many of those aspects can be addressed simultaneously or need to be balanced. For instance, while a fast curing speed is often required for fast processing, the generally used radical curing mechanism results in an oxygen sensitivity, so the presence of oxygen will prevent the cross-linking reaction to occur. On the other hand, processing requirements for the imprinting step, such as low forces during the separation step may contradict final application needs such as hydrophilicity for micro- and nanofluidic behavior. Our goal was to develop a material formulation that shows both, a reduced oxygen sensitivity and an increased hydrophilicity as well as the possibility of post-functionalization by having active surface groups.

To accomplish this, in collaboration with our colleagues at Micro Resist Technology GmbH (Berlin, Germany), we used a variety of formulations, as summarized in Table 1. We started off with commercial monomers. We used an aliphatic polyurethane acrylate (PUA) as the base resin (formulation 1), since it is widely used within the community and easy to work with. In addition, it gives overall good hardness and adhesion towards polymeric surfaces. In order to introduce the hydrophilicity, two different trifunctional monomers were used, as detailed in Table 1. Each one introduced flexibility and hygroscopicity to the overall formulation that needed to be balanced with the amount of PUA used. In particular, reducing the oxygen sensitivity and introducing free thiol groups was accomplished by using hydrophilic trifunctional thiol monomers. In order to reduce the inevitable dark reaction occurring in thiol-ene based formulations, proprietary stabilizers were used.

With those commercially available monomers, first the hydrophilicity was adjusted (formulations 2-5) and later on, the oxygen sensitivity using the thiol-ene analog (formulation 6).

Formulation 6 showed optimal material properties, including the targeted presence of free thiol groups on the surface. The formulation is easily patternable by UV-NIL. The cured material is transparent, and the surface hydrophilic. The presence of the thiol groups was confirmed by Raman spectroscopy, and their functionalization by click chemistry as shown later in this article.

**Table 1.** Summary of the used monomers for prototyping a hydrophilic UV-curable formulation with post-functional properties.

| Employed formulation components | Main impact on cured material | Formulation name | | | | | |
|---|---|---|---|---|---|---|---|
| | | 1 | 2 | 3 | 4 | 5 | 6 |
| | | Monomer concentration range (%) | | | | | |
| Aliphatic polyurethane acrylate[a] (PUA) | Hydrophobicity, High cohesion, hardness, and glass transition temperature | 100 | 90 | 70 | 50 | 0 | 25 |
| 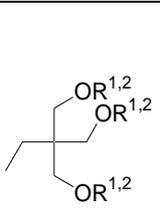 R$^1$ = 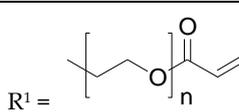 | High flexibility, high hydrophilicity | 0 | 10 | 30 | 50 | 100 | 30 |
| R$^2$ = 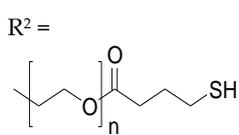 | Flexibility, oxygen insensitivity, post-functionalization of thiol groups | 0 | 0 | 0 | 0 | 0 | 40 |
| Stabilizers | | | | | | | >1 |
| Photoinitiator | | 2–5 | 2–5 | 2–5 | 2–5 | 2–5 | 2–5 |

[a] Structure not defined by material supplier

*3.2. Hydrophilicity and Hygroscopicity of Base Formulation*

The hydrophilicity of the polymer surface is determined by the concentration and chemistry of the surface groups. Since water is a polar liquid, surfaces with higher concentrations of polar groups will have lower contact angles than those with non-polar groups. Thus, to make the polymer more or less hydrophilic, we have tuned the ratio between the hydrophobic polyurethane acrylate (PUA) and the trifunctional hydrophilic acrylate (see Table 1, formulations 1-5). In addition, the molecular length of these monomers can be changed for fine-tuning, by increasing the amount of ethylene oxide (EO) moieties per monomer in the synthesis.

We investigated the hydrophilicity of several formulations. For this, we measured the advancing dynamic water contact angle (WCA) on the resulting materials after UV curing under different conditions, as shown in Figure 2. In particular, we varied the chemical composition of the formulations: the weight percentage of the hydrophilic trifunctional monomer and the length of the polyglycol groups, indicated by the relative ethylene oxide (EO) amount in the synthesis of the monomer.

Figure 2, red circles, shows the WCA for the formulations shown in Table 1. To our surprise, there was very little change in the hydrophilicity of the different surfaces when curing the material under inert, non-polar conditions. We used a $CO_2$ atmosphere, but $N_2$ or Argon should work similarly. Neither changing the monomer, nor the ratio changed the resulting hydrophilicity significantly. WCAs of around 70-75° were observed, even though up to 50% of hydrophilic monomer

was used, as indicated by the red squares in the graph in Figure 2. Only when using 100% of the trifunctional hydrophilic monomer, a WCA of 60° was reached (formulation 5).

On the other hand, when curing under hydrophilic (polar) conditions (using a non-structured OrmoComp® stamp, plasma activated; Micro Resist Technology, Berlin, Germany) the overall WCA dropped to roughly 54°, almost indifferent of the hydrophilic content of the formulation. The red triangles in the graph in Figure 2 show the WCA of formulations 2, 3 and 4, cured under a polar surface. Varying the length of the ethylene glycol chain while maintaining the monomer concentration at 50%, the WCA dropped down to 47°. This can be seen also in the graph, where the black, red, green and blue markers correspond to different variations of formulation 4, with increasing amount of EO moieties per monomer (i.e., with different lengths of the polyglycol groups).

One possible explanation for this strong dependence on the curing conditions could be related to the free energy diffusion of the monomers before and during curing. In a non-polar surrounding, the hydrophobic (i.e., non-polar) parts of the formulation will arrange at the surface and will be fixated there upon irradiation. In the second case, the behavior is the opposite, and the polar parts will move to the surface when the uncured resist is in contact with the polar stamp. This could also explain why there is small variation in the contact angle when comparing different monomer compositions.

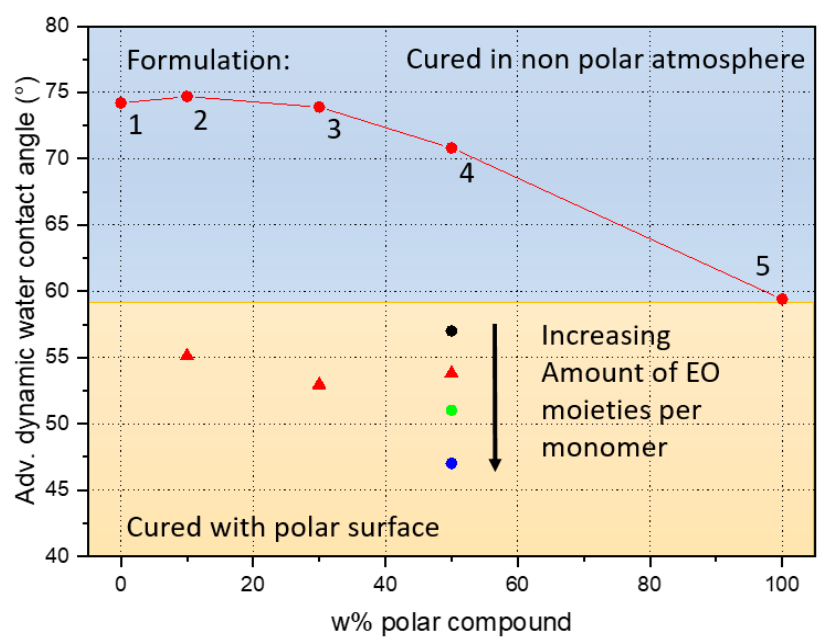

**Figure 2.** Evaluation of the effect of the chemical composition of the thiol polymer prototypes on its hydrophilicity, measured by the advancing dynamic contact angle. Formulations with a varying polar components content are shown in red dots for a 10%, 30% and 50%, all with WCA below 57°, when cured in contact with a polar surface (i.e., using a non-structured Ormocomp stamp).

The hydrophilicity does not come without the drawback of a certain amount of water absorption into the polymer (hygroscopicity). Since the water absorption might lead to swelling, the hygroscopicity should be accounted for if the feature size is critical. Formulations 1-5 (see table 1) have been cured and the bulk layers (roughly 500 µm thick) were tested for water absorption. Depending on the amount of hydrophilic monomer, the water absorption varied between 1 to 30 w% depending on the length of the ethylene glycol moieties and the amount of polar monomer added to the formulation.

## 4. Thiol-Based Polymer

The results from those tests were used for defining the optimal material (formulation 6), which balances both the hydrophilicity and hygroscopicity, offering a good compromise: the cured material has a water uptake of less than 2%, while still having enough excess thiol groups on the surface for post functionalization. This material has a dynamic contact angle of 40°, which we have measured on samples cured without and with a stamp, functionalized with a monolayer of fluorosilanes.

*4.1. Hydrophilicity and Hygroscopicity of Thiol Formulation*

We have observed a non-steady behavior of water sessile drops on the surface of the cured film. Due to the interaction of the water with the surface a decrease in the observed contact angle can be seen. As the cured polymer shows a certain degree of hygroscopicity, it absorbs water and the polymer/water interface changes over time. The time dependent change of the contact angle of a water drop on untreated, pristine thiol polymer is shown in Figure 3 (black line). This pristine sample, fabricated by spin coating and curing with 2000 mJ/cm$^2$ in an EVG 501 shows an initial contact angle of 47°, which decreases in two minutes by 7°.

A surface activation using UV-Ozone or oxygen plasma oxidizes the organic groups at the surface, leaving more hydroxyl (–OH) groups exposed, increasing the polymer's hydrophilicity. Figure 3 (red line) shows the change of the WCA for a polymer layer treated in a UV-Ozone plasma for four minutes. The treatment reduces the contact angle by 10° with respect to the untreated, pristine sample. Its time development is very similar to the pristine sample, indicating that the UV-Ozone results in only a chemical surface modification.

A similar measurement, obtained after treating a sample in oxygen plasma (Technics Plasma 100-E) for four minutes is shown in Figure 3 (blue line). The initial wetting resulted in a higher contact angle than that of the pristine sample, but dropped within ten seconds below the value obtained for the UV-Ozone treated sample. The WCA becomes constant after a couple of minutes with a value of 23°.

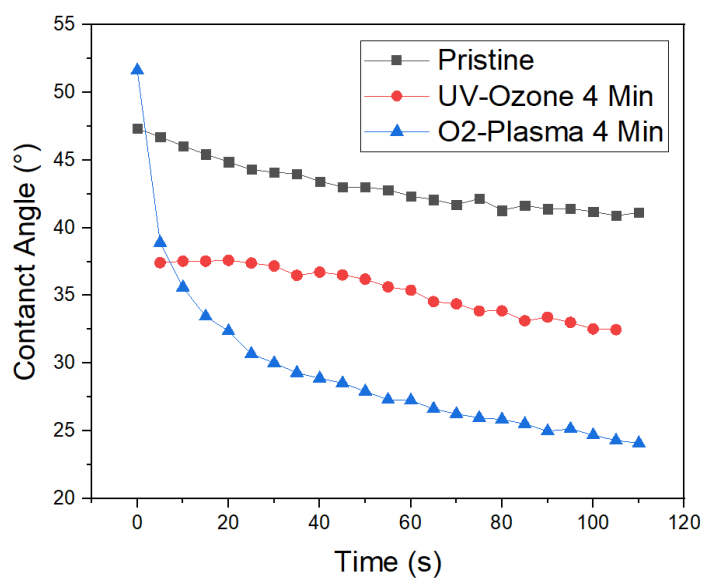

**Figure 3.** Contact angle measured on an untreated, pristine cured polymer layer (black line), a sample after a four-minute UV-Ozone treatment (red line) and a sample after a four-minute O$_2$ plasma activation (blue line). The WCA was measured every five seconds, and the graph shows its evolution along time for the three different samples.

The increased hydrophilicity after UV-ozone and O$_2$ plasma are stable for few hours—we have measured no change in the WCA two hours after the treatment. But the effect of the treatment is not permanent. We have observed that, after few days, the measured contact angle for the UV-ozone treated samples increases and the initial value is recovered after one week, when storing the samples at room temperature and ambient conditions. The regain of the surface properties can be attributed to a rearrangement of functional groups at the surface. This entropy-driven process will result in the diffusion and rearrangement of the short polymeric chains at the surface: the polar groups will be oriented towards the bulk and the more hydrophobic groups to the surface, realizing an equilibrium matching the atmosphere [26]. The $O_2$ plasma changes more drastically the contact angle, and leads to a more durable change, since it also introduces a nanoscopic surface roughness. And, in hydrophilic surfaces, higher roughness leads to lower WCA [27,28].

Since the $O_2$ plasma etches the polymer (in our case, <5 nm/min), it should be used with precaution, especially when dealing with features with small dimensions, or when the roughness of the surface is a critical parameter.

*4.2. Presence of Thiol Groups*

4.2.1. Raman Spectroscopy

When curing (meth)acrylates in the presence of thiols, two competitive reactions occur: the homopolymerization of the acrylates and the photo-initiated thiol-ene reaction. Since both reactions have different mechanisms (chain growth vs. step growth [22,23]) which take place on the same time-scale, an homogenous distribution of thiol groups in the bulk and on the surface after polymerization can be expected. In order to confirm the presence of thiol groups on the surface of the cured polymer, and also inside the bulk material, we performed Raman spectroscopy on the cured films to check for the vibrational peak corresponding to the thiol group (R-SH) at 2575 cm$^{-1}$.

Figure 4 compares the Raman scattering spectrum of different polymers. The red line in the graph corresponds to a thiol-containing polymer (Table 1, formulation 6), and the black line to an exemplary organic, thiol-free polymer with similar formulation. The R–SH vibrational peak is apparent only in the thiol polymer measurements, confirming the presence of thiol groups just in the thiol polymer prototype.

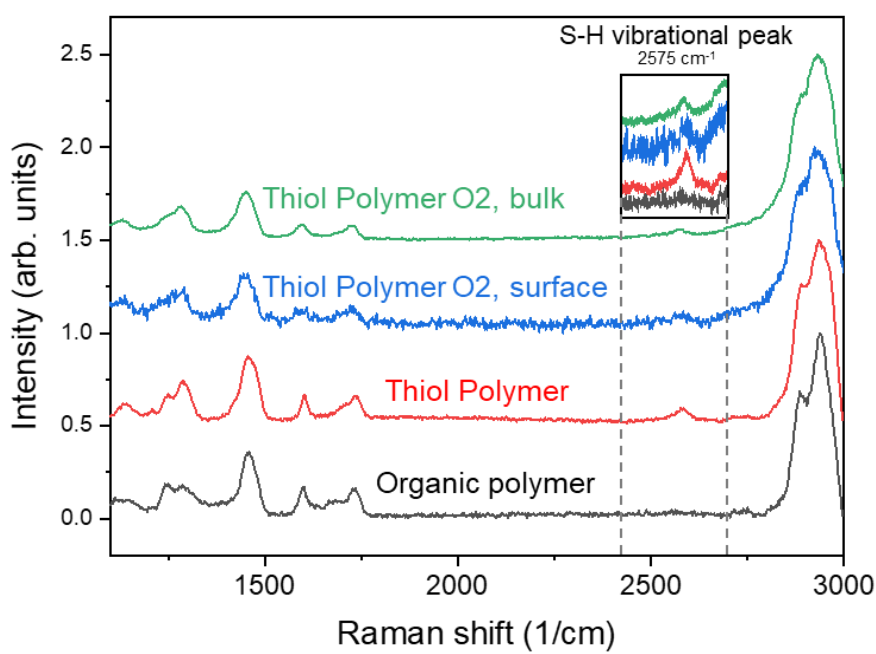

**Figure 4.** Raman scattering of the polymer samples. The graph shows the signal obtained on a non-thiol polymer ("Biopolymer", an organic polymer, used as control) (black line). The red line shows the signal from a thiol polymer film. The blue and green signals correspond to a film etched using an $O_2$ plasma, and measured at the surface and the bulk respectively. The inset shows a detail of the peak at 2575cm$^{-1}$, which corresponds to the R–SH vibrational mode of the thiol groups.

We also selectively etched the thiol-polymer with an $O_2$ Plasma (Technics Plasma 100-E) for four minutes and measured the Raman scattering spectrum again (Figure 4, blue line). The etching process and surface modification after the oxygen plasma treatment does not lead to a significant difference in the signal. Furthermore, by changing the focal point of the laser, we can obtain the signal from the bulk material. Figure 4 (green line) shows the signal obtained from the bulk material, where the 2575 cm$^{-1}$ peak is still present, confirming that there are thiol groups also in the bulk material. As mentioned before, the thiol groups are part of the cured, cross-linked monomers, and thus, are an intrinsic part of the bulk polymer. After polymer etching, the new groups from the newly created surface will now become superficial.

4.2.2. Surface Functionalization

As it has been described previously in the literature [20], excess thiol groups can be used to functionalize surfaces. This allows one, for example, to alter their wetting behavior. Or, more interestingly, to make them active and selective to certain chemical groups, for example to covalently attach anchor groups that selectively bind to specific biomolecules. As a proof of concept, we observed the change in the surface properties after performing a Michael addition on the cured thiol polymer, as sketched in Figure 5a–c. For this, a mixture of a fluorinated acrylate ($C_{12}H_5F_{15}O_2$) and trimethylamine (1 mmol solution) was incubated on the cured polymer film for 24h at room temperature (Figure 5b). The solution was then washed with isopropanol and dried (Figure 5 (c)). During the incubation, the free thiol groups reacted with the acrylates, as sketched in (c), which resulted in an increase in the hydrophobicity. Since this reaction is not radical driven, but a nucleophilic attack, no side reactions or grafting of the polymer chain are expected. We have seen that the water contact angle of the polymer surface after curing was 40°, as shown in Figure 5 (d), and that it increased up to 77° after the functionalization, as shown in (e). Full coverage with a self-assembled monolayer of fluorosilanes typically results in WCA of >100° [29]. Thus, since we observe a WCA of 77°, we assume that we don't have a full coverage. We estimated it in the range of 40–60%, which is probably reflecting the amount of thiol groups at the surface. To increase the coverage, a larger amount of surface thiol groups should be present at the surface; for this, a different formulation should be used, at the expense of different material properties.

This experiment proofs that the thiol groups at the surface are active and reactive to chemical modifications, opening the way to direct, simple surface modifications using wet chemistry, since the process can be performed with other "clickable" monomers.

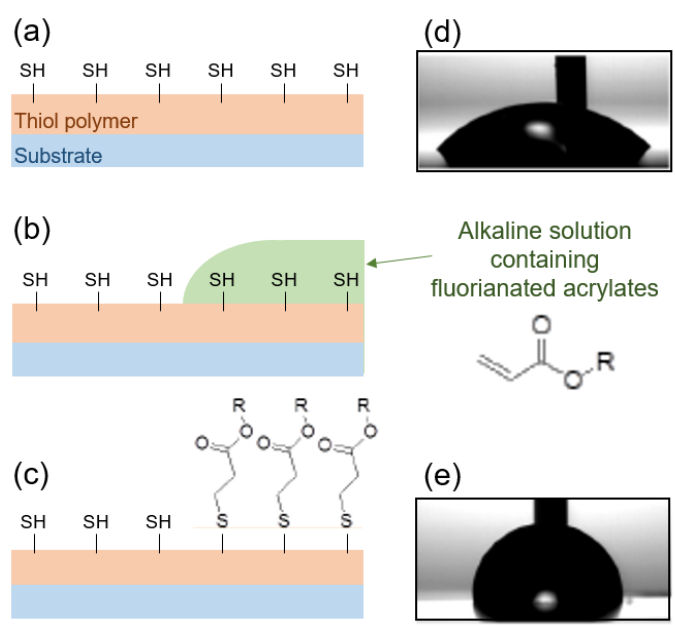

**Figure 5.** Surface chemistry on the polymer using the thiol functional groups. An imprinted thiol polymer sample (**a**), which has surface-active thiol groups, is incubated with a solution of fluorinated acrylate and trimethylamine (**b**) for 24h. The molecules selectively react with the thiol functional groups in a Michael addition (**c**). The contact angle of the untreated sample is 40° (**d**), while after the chemical surface modification, it increases to 77° (**e**).

*4.3. Material Properties and Patterning*

4.3.1. General Properties

The cured polymer is transparent to visible light and has a slight yellowish color, resulting from the used photoinitiator, when cured in thick layers of several micrometers. The transmittance is 99.5% for visible light ($\lambda \geq 380$ nm), measured on a 1.6 µm thick layer of cured polymer on a glass slide. The transmission spectrum for the visible range can be seen in Figure S1 in the supplementary material. The polymer, owing to its acrylate and thiol-ene chemistry, shows typical behavior against solvents and acids. It is resistant to isopropanol, but not to acetone. It shows a high optical transparency (>99%) in the visible range with a refractive index of 1.486. Its intrinsic hydrophilicity (water contact angle 40°) is low enough for spontaneous microfluidic behavior in open channel microfluidics and top surface functional groups (R–SH, and R–OH) have been reported. These and other general properties of the investigated prototype liquid and cured resist are given in Table 2.

**Table 2.** General properties of the polymer before and after curing.

| **Uncured Material** | |
|---|---|
| **Viscosity** | 330 mPas |
| Mold compatibility | Silicon, Glass, Ormostamp |
| Curing atmosphere | Oxygen free |
| Curing dose | >2000 mJ/cm$^2$ |
| **Cured material** | |
| Refractive index | 1.486 |
| Transparency ($\lambda \geq 350$ nm) | >95% |
| Transparency ($\lambda \geq 380$ nm) | >99% |
| Surface functional groups | SH, OH |
| Contact Angle | 40° |
| Layer thickness (spin coating, 3000 rpm) | 5.0 µm |
| Layer thickness (spin coating, 6000 rpm) | 1.9 µm |
| Layer thickness (drop casting) | 20–70 µm |

4.3.2. Patterning

The patterning of the UV-resist is sketched in Figure 1a–c: the material is drop casted or spin coated on the substrate (a), then, a hard stamp is brought into contact, and the polymer is cured under UV light (b); the stamp is separated from the substrate, and the pattern is replicated on the polymer (c). The typical thickness of the drop-casted imprinted layers is around 70 µm. Drop casting is a manual process, so the droplet volume and changes in the viscosity (associated to temperature fluctuations) cause a variation in the final layer thicknesses. Reproducible and controlled layer thicknesses are achieved by spin coating. Film thickness of 1.9 and 5.0 µm were obtained for spin speeds of 3000 and 6000 rpm, respectively.

The developed material prototype can be processed both by spin coating or drop casting. Different materials, like silicon, glass or polymers (e.g. PMMA and polycarbonate, PC) are suitable substrates. The adhesion to the substrate can be improved by cleaning and activation by UV-Ozone or oxygen plasma; an adhesion promoter is recommended for inorganic substrates. Since it is solvent free, no pre-baking step is necessary, simplifying the processing and improving the throughput and compatibility to different substrates. The polymer is cross-linked by exposing it to monochromatic or broadband UV light. A minimum curing dose of 2000 mJ/cm² is recommended. This leads to a high transparency and hardness.

Gas-impermeable stamps should be used (like glass or silicon), since the curing chemistry is sensitive to oxygen. Thus, PDMS stamps should be avoided. All imprints shown in this work have been performed with hard, non-gas permeable stamps, made of silicon or cured Ormostamp®, to avoid any interfering oxygen at the surface or diffusion into the polymerizing liquid. Further high intensity UV irradiation was used to overcome the effect of possible residual oxygen. The low viscosity of the uncured material enables imprinting at room temperature with low pressure: the weight of the stamp or the sample is usually sufficient to start filling the cavities, which is then followed by capillary forces that result in a complete filling with the liquid resist within seconds. In order to decrease the releasing forces during the separation of the stamp and the cured resist, the stamps were coated with an anti-sticking layer of fluorosilanes by chemical vapor deposition [29].

In this work, we have used an EVG Imprint tool (EVG 501 customized for UV and thermal imprinting) for sample imprinting. The tool allows for chamber evacuation before the NIL process, to remove oxygen from the liquid polymer. The chamber is evacuated to 0.1 mbar for 20 seconds to degas the polymer and then purged again with air. The exposure is done with a pulsed, broadband UV Xenon lamp (190 nm cutoff) with 105 mW/cm² for 19 seconds, resulting in a dose of 2000 mJ/cm². Other tests with a non-pulsed, low energy density, monochromatic UV LED lamp (365 nm) and a custom-made flow box to locally create an inert gas atmosphere worked as well.

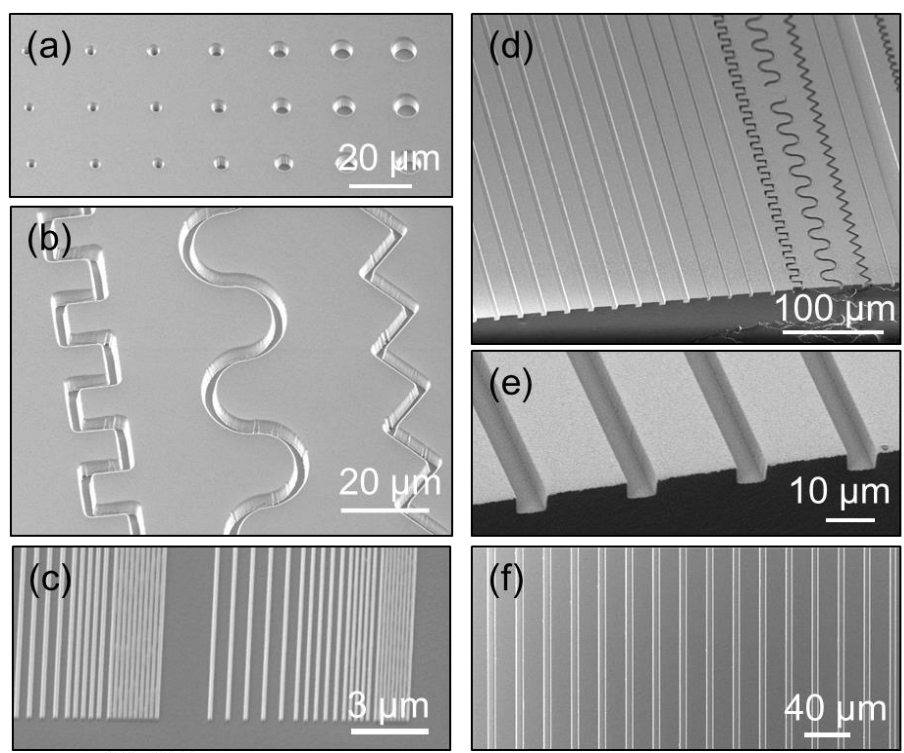

**Figure 6.** SEM images of micro- and nano-imprinted structures are shown. The structures vary from dots, lines, meander structures, and rely only on the quality of the structures in the stamp. Micro

structures with different shapes and configurations are shown in (**a**,**b**,**d**,**e**,**f**) (dots, meandering lines, straight lines). And also lines with nanometric dimensions, down to 75 nm wide, as shown in (**c**).

Figure 6 shows scanning electron microscope (SEM) images of thiol-polymer imprinted structures with lateral dimensions in the micro- and nanoscale. Dots with diameters ranging from 1.2 to 11 µm can be seen in (a) and meandering lines, 2 µm wide and 4.8 µm deep in (b). Nanostructures and lines with gradually smaller widths from 200 to 75 nm are shown in (c). (d) shows tilted SEM images (30°) of a combination of a 20 µm wide line and several straight and meander structures, 5 µm wide, which cross section can be seen in (e). The top view of these identical lines can be seen in (f). The smallest feature sizes obtained were 75 nm wide and 200 nm height with a pitch of 150 nm. It is noteworthy that the shown resolution of the nanostructures is mainly limited by the stamp; imprinting of smaller features should be possible. The developed thiol-polymer prototype allows for simple, easy and reproducible combination of structures with large and small dimensions without problems.

### 4.3.3. Etching

The thiol polymer prototype can be structured and removed by reactive ion etching (RIE). Standard pattering techniques leave a residual layer after imprinting. In order to remove this layer for etching and pattern transfer, RIE has been proven to be a suitable method. We have used an ICP-RIE system (SI 500 RIE, from Sentech) to test the etching properties for standard pure $O_2$ and $SF_6$-based recipes, typically used for etching micro and nano structures into polymers or silicon, respectively. The results for the etching rate for the polymer for the two different processes are shown in Table 3. The etching rate of 0.4 µm/min obtained for the $O_2$ plasma is comparable to that for other typical organic polymers etched under similar conditions. The resulting surface is very smooth, as can be seen in Figure S2a in the supplementary information.

**Table 3.** Etching of the thiol polymer by RIE. Two typical RIE recipes used to remove or structure polymers ($O_2$ etch) and silicon (Si etch) were evaluated.

| Process | ICP Power | HF Power | Gas | Pressure (Pa) | Flow (sccm) | Temp. (°C) | Etch Rate (µm/min) |
|---|---|---|---|---|---|---|---|
| $O_2$ etch | 150 W | 50 W | $O_2$ | 0.5 | 30 | 21 | 0.4 |
| Si etch (anisotropic) | 400 W | 15 W | $O_2$, $SF_6$, $C_4F_8$ | 1.0 | 5, 50, 70 | 3 | 0.2 (Si 0.6) |

Polymers are often used as masks for etching or are part of more complex devices which undergo further processing. For this, we have tested the polymer also against recipes commonly used to etch silicon. We used a typical recipe, where $SF_6$, $C_4F_8$ and $O_2$ are mixed (details shown in Table 3). With this process, the polymer is etched at 0.2 µm/min and silicon at 0.6 µm/min. This gives a selectivity of 1:3. This recipe leads to rougher surface even for short etching times (100 s), as can be seen in Figure S2 (b) in the supplementary information. We also tested the polymer in an $O_2$ $SF_6$ RIE (without the passivant, $C_4F_8$), but obtained worse selectivity against silicon, and rougher surfaces (see Figure S2c,d).

It should be noted that in our reactors chamber, which contains aluminum, we observe some times Al sputtering on the surface during some of the RIE processes—especially those with high chamber pressures; small clusters of aluminium are deposited on the polymer surface and act as a mask for the etching, resulting in grass-like polymer structures (see Figure S2d in the supplementary material). This "grass effect" has been reported in the literature and is not unique to this polymer [30].

### *4.4. Application: Electrical Measurements on Graphene*

One further application to exploit the intrinsic properties of the newly developed material lays in the field of graphene field effect transistors (GFET). Graphene is a two dimensional, highly conductive material. Since its conductivity is strongly dependent on the purity of the material and on the surface properties, even minor amounts of adhered molecules will result in doping and thereby a strong change in the properties of the GFET. One critical adhering molecule can be water from the humidity in the air. In order to obtain reliable data under ambient conditions, water needs to be removed. Thus, electrical measurements of GFETS are typically performed at ultra-high vacuum conditions. We have found that placing a cured layer of the polymer developed here (formulation 6) directly on top of the GFET results in a decreased doping, as the resist will absorb the moisture due to its hygroscopicity.

We compared the conductivity of a graphene field effect transistor patterned on a silicon oxide substrate with and without the cured thiol polymer adhered on top, as shown in the sketch in Figure 7a and in the image of an actual device in Figure 7b, using a back gate to control the charge carrier type and concentration. This is a standard device configuration to quantify doping in graphene [31]. By measuring the resistance as a function of the gate voltage, as shown in the graph in Figure 7c, the maximum resistivity indicates the voltage needed to empty the conductance band and completely fill the valence band of the graphene; this value is the so-called charge neutrality point (CNP). Intrinsically, the CNP of graphene lies at zero volts and shifts due to doping (for example due to adhered molecules or humidity on the surface). Figure 7c, green line, shows the resistance for a sweep in the gate voltage for the uncovered graphene transistor at ambient conditions (room temperature, and a humidity of 40–50%). The curve can be better seen in the inset. The red curves in the graph show the signal of the same GFET, patterned using the same graphene flake, where it was locally covered by the polymer (as sketched and shown in the drawing and image in Figure 7a,b). The presence of the polymer leads to a very strong shift of the CNP from higher positive gate voltages towards the zero bias region. We confirmed this effect in two additional graphene devices.

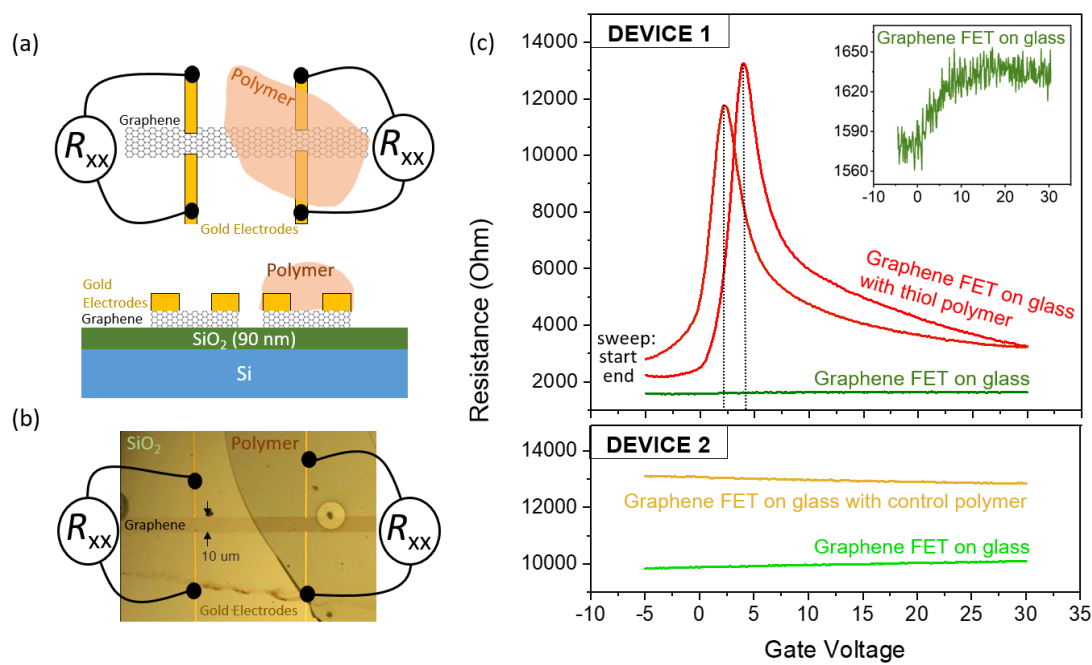

**Figure 7.** (**a**) Graphene field effect transistor (GFET). A graphene flake is structured on a SiO$_2$/Si substrate, and gold electrodes patterned on top, to act as electrical contacts for measurements. Part of the GFET is locally covered with a cured piece of thiol-polymer, as sketched in the top and side views. The p-doped silicon substrate acts as a back gate, which is electrically isolated from the graphene Hall bar by a layer of SiO$_2$ (bottom panel). The upper part of the Hall bar is covered by polymer whereas the lower part remains exposed. (**b**) Photo of an actual device, where the area on the right side is darker due to polymer deposited on top. (**c**) Measurement of the two-terminal resistances of the GFET device as a function of a gate voltage. The upper graph (Device 1) shows the results for one of the measured devices. The green line corresponds to the graphene "at-air". The red lines correspond to the measurement on the graphene covered with the thiol-polymer, for a sweep from −5 V to 30 V, and from 30 V to −5 V, as marked in the graph. The lower graph (Device 2) shows the results of a similar experiment performed in another different graphene FET device, where instead of the thiol polymer, a layer of a non-hygroscopic polymer (Ormostamp) is used to cover the graphene, as control experiment.

Since the measurements were done at ambient conditions in clean GFETs, just water is expected to be adhered to the surface of the graphene flake. Thus, the adsorbed water molecules dope the graphene (p-doping), shifting the CNP towards higher gate voltages [32]. As a control experiment, we covered a different GFET with a non-hygroscopic polymer (Ormostamp®). These measurements on "Device 2" are shown in the bottom panel of Figure 7c. The resistance as a function of the gate voltage measured for the uncovered GFET Device 2 "at air" is shown in the green line and shows a similar behavior as Device 1. The base resistance differs since the intrinsic doping can vary between different devices. As we put a layer of cured Ormostamp® on the graphene, a change in the signal is observed, as can be seen in the orange line. However, the hygroscopic effect of the thiol polymer on graphene as seen for Device 1 is not reproduced with this other polymer.

These results underline the relevance of the hygroscopic properties of the thiol polymer and its possible application for example for graphene based electronics, which could function at ambient conditions [33]. In addition, this also enables the operation close to the CNP. Changes in the charge carrier density around the CNP result in the strongest changes in conductivity, enhancing the graphene's sensitivity for small perturbations, which is highly desirable in detectors.

## 5. Conclusions

A polymer prototype containing thiol functional groups and an intrinsic hydrophilicity was presented. The polymer is UV curable, and can be patterned at the micro and nanoscale by UV nanoimprint lithography. The thiol groups present at the surface can be directly used for chemical functionalization.

The presented formulation uses PUA as base resin, and two different trifunctional monomers. To adjust the hydrophilicity of the product, the contact angle of cured material made of different mixtures was measured after curing the material. It was observed that the percentage in weight of the hydrophilic components played a minor role, and that mainly the curing conditions determined the final contact angle.

A mixture of 25% PUA, 35% trifunctional monomer and 40% trifunctional thiol monomer was proposed as the optimal formulation. This material has a contact angle of 40° and a small, intrinsic hygroscopicity. The polymer is highly transparent in the visible range (T > 99%) and has a refractive index of n = 1.486. Structures with lateral dimensions spanning from several microns down to 75 nm were imprinted by UV-NIL. The dry etching (RIE) of the polymer was investigated for $O_2$ and $O_2 + C_4F_8 + SF_6$ plasmas, and the selectivity to silicon was shown to be 1:3.

The presence of the thiol groups in the optimal formulation was confirmed by Raman spectroscopy. The peak corresponding to the vibrational mode corresponding to the thiol group at 2575 $cm^{-1}$ was present in the cured polymer samples, on the surface and also in the bulk material, confirming that the thiol groups are an intrinsic part of the polymer. Even after $O_2$ plasma etching, the Raman peak corresponding the thiol groups is present, confirming the compatibility of the polymer with standard $O_2$ plasma processing.

To further prove the presence of thiol groups at the surface, and to show one example of the potential of the polymer for applications requiring surface chemistry (e.g., "click chemistry"), a Michael addition was performed. A solution of fluorinated acrylate in trimethylamine was incubated on a flat, cured polymer surface, resulting in a significant increase in the water contact angle as a result of the change in the surface chemistry, showing a surface coverage in the range of 40 to 60%.

As another example of application, we have shown that the cured thiol polymer has a strong effect on the electrical properties of graphene. We measured the CNP of a GFET device at ambient conditions and showed a large shift towards zero when the graphene was coated with a thin layer of polymer, as compared to a non-coated one. Since the polymer is hygroscopic, it absorbs the residual layer of water (originating from ambient humidity) from the graphene surface, leading to electrical properties of the GFET similar to those which can only be obtained under vacuum conditions when the devices are not coated with the polymer layer.

The work presented in this paper shows the potential and versatility of the proposed polymer, which is easy to process and pattern at the micro and nanoscale, and which allows for a variety of applications yet to be explored. Our preliminary results pave the way for using the polymer for click chemistry processes and opens an application window for simple imprinting of complex nanostructures and consecutive easy chemical surface modifications.

**Supplementary Materials:** The following are available online at www.mdpi.com/xxx/s1, Figure S1: Polymer transparency and Figure S2: Surface roughness.

**Author Contributions:** Conceptualization, methodology, I.F.-C. and M.M.; Validation and investigation M.M., L.T., R.N.; writing – original draft preparation M.M and I.F.C; writing – review and editing M.M., I.F.-C., L.T., R.N.; funding acquisition, I.F.-C. All authors have read and agreed to the published version of the manuscript.

**Funding:** This project has received funding from the European Research Council (ERC) under the European Union's Horizon 2020 research and innovation program (grant agreement No 714073). This research was also partially funded from the European Union's Horizon 2020 research and innovation program under grant agreement No 646260.

**Acknowledgments:** Authors would like to thank the team at micro resist technology GmbH in general, and Mirko Lohse and Manuel Thesen in particular, for the help with the prototype development and the support with the measurements

**Conflicts of Interest:** The authors declare no conflict of interest